\documentclass[pre,twocolumn,11pt]{revtex4-1}
\usepackage{bm}%bold math
\usepackage{graphicx}
\usepackage{amsmath}
\usepackage{amssymb}
\usepackage{setspace}
\usepackage[english]{babel}
\begin{document}

%\preprint{}

\title{A Simple Model for Long-Range Interacting Pendula}



\author{Owen Myers}
\affiliation{Materials Science Program, University of Vermont, Burlington, Vermont 05405, USA.}
\affiliation{Department of Physics, University of Vermont, Burlington, Vermont 05405, USA.}
%\email{oweenm@gmail.com \\ jwu@uvm.edu}
\email{oweenm@gmail.com}
\author{Adrian Del Maestro}
\affiliation{Materials Science Program, University of Vermont, Burlington, Vermont 05405, USA.}
\affiliation{Department of Physics, University of Vermont, Burlington, Vermont 05405, USA.}
\author{Junru Wu}
\affiliation{Materials Science Program, University of Vermont, Burlington, Vermont 05405, USA.}
\affiliation{Department of Physics, University of Vermont, Burlington, Vermont 05405, USA.}
\author{Jeffrey S. Marshall}
\affiliation{School of Engineering, University of Vermont, Burlington, Vermont 05405, USA.}

%\thanks{}

\date{\today}

\begin{abstract}
We show that the Hamiltonian mean field (HMF) model describes the equilibrium behavior of a system
of long pendula with flat bobs that are coupled through long-range interactions (charged or self
gravitating). We solve for the canonical partition function in the coordinate frame of the pendula
angles. The Hamiltonian in the angles coordinate frame looks similar to the form of the HMF model
but with the inclusion of an index dependent phase in the interaction term. We also show interesting
non-equilibrium behavior of the pendula angles, namely that a quasistationary clustered state can
exist when pendula angles are initially ordered by their index.  
\end{abstract}

\pacs{}% insert suggested PACS numbers in braces on next line

\maketitle 

% If in two-column mode, this environment will change to single-column format so that long equations can be displayed. 
%\begin{widetext}
%$$\mbox{put long equation here}$$
%\end{widetext}

% \begin{table}
% \end{tabular}
% \end{table}

\section{Introduction}
Systems with long-range interactions are a source of unique problems in the field of statistical
mechanics and thermodynamics. This is due to several properties of long-range systems which fall
outside of the conditions normally needing to be satisfied when applying the methodologies of
thermodynamics. Simply from the words ``long-range'' the first infringement can be deduced, that
long-range systems are not additive.  If two systems with short-range interactions are brought
together to form a larger system then the energy difference between the conglomerate system and the
sum of its constituents is the new potential energy from the boundary between them. In the
thermodynamic limit, the potential energy of the boundary is small compared to the bulk and can be
neglected, making short-range systems additive.  In the case of long-range interactions, one
particle will feel a significant potential created by every other particle, so the additional
potential energy of two systems added together does not scale as the boundary but in a more
complicated way that depends on the specific nature of the interactions \cite{dynam_therm_intro}.
Directly related to the lack of additivity is the fact that systems with long-range interactions are
not extensive because their energy diverges in the thermodynamic limit \cite{kac1963}.  Although
these characteristics compel cautious use of the usual tools of statistical mechanics, they are also
the source of many interesting dynamical and statistical features.  Depending on the system of
interest, such features include canonical and microcanonical ensemble inequivalence and related
negative specific heat \cite{sire2002}, quasistationary states (different than metastable states
which lie on local extrema of equilibrium potentials) whose lifetimes increase with the number of
particles \cite{antoniazzi2007}, an interesting dependence of the largest Lyapunov exponent on particle
number in a long-range Fermi-Pasta-Ulam model \cite{christodoulidi2014}, and spontaneous creation of macroscopic structures in
non-equilibrium states \cite{antoni1995}.  In some cases, long-range interactions can greatly
simplify problems. For instance, mean field models depend on one of two premises: (i) interactions
are short-range but the system is embedded in a space of infinite dimension so that all bodies in
the system are nearest neighbors, or (ii) interactions are infinitely long.

For some time, the primary motivation for the study of long-range interactions was to understand galaxies, galaxy
clusters and the general thermodynamic properties of self-gravitating systems. Aside from mean field
models, interest has further built since the observation of modified scattering lengths in
Bose-Einstein Condensates (BEC) through the use of Feshbach resonances \cite{inouye1998}.  Using
this technique, a BEC can be made to be almost non-interacting by tuning the scattering length to
zero.  One could even tune the scattering length to a negative value, making the BEC collapse. More
recently, O'dell et al. \cite{odell2000} has shown that it may be possible to produce an attractive
$1/r$ potential between atoms in a BEC by applying an ``extremely off resonant'' electromagnetic
field. This has opened the possibility of creating table-top methods which physically model
aspects of cosmological behavior on a laboratory scale, as well as the possible development of entirely new
dynamics in BEC. 

\begin{figure}[h]
    \centering
    \includegraphics[width=8.0cm]{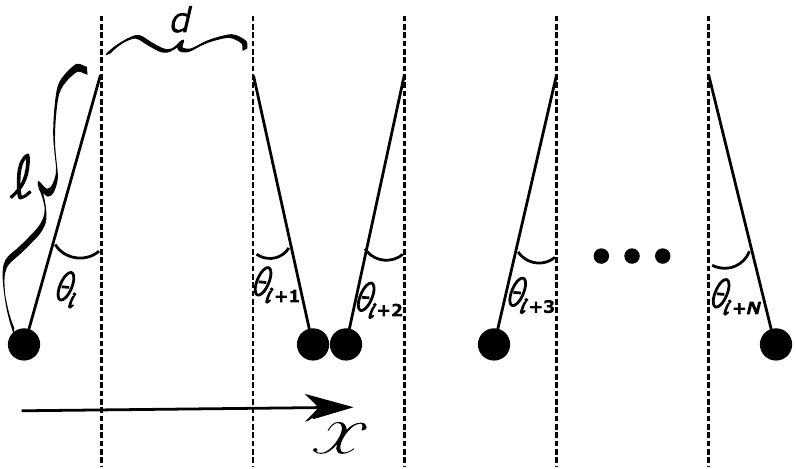}
    \caption{$N$ pendulum system with parallel planes of rotation. The $i^{\mathrm{th}}$ pendulum angle at
    some time $t$ is $\theta_i(t)$. \label{fig:system}}
\end{figure}

The challenges in understanding long-range systems drive the development of solvable models that
could help better explain some of the aforementioned phenomena. Campa et al. \cite{campa2009} have recently published
a collection of important solvable models. One particularly significant model,
which is important to this work, is the Hamiltonian Mean Field (HMF) $XY$ spin model
\cite{antoni1995}, often written in the form

\begin{equation}
    \label{eqn:HMF}
    H = \sum_{i=1}^{N} \frac{p_i^2}{2} + \frac{\gamma}{2N}\sum_{i,j=1}^{N} 
    \left[
        1- \cos{(\theta_i-\theta_j)}
    \right]
    ,
\end{equation}

\noindent
where $\theta_i$ is the angular position of the $i^{\mathrm{th}}$ particle (spin), as shown in
Fig.~\ref{fig:system}, and $p_i$ is its
conjugate angular momentum. The HMF model is
generally used to describe two different classes of systems: 1) a mean field $XY$ classical spin
model, and 2) a one dimensional periodic system of itinerant
particles with long-range interactions. Though the connection between the HMF model and the second class of systems mentioned could be
thought of as contrived given the simplifications under which the model is realized, it has been
shown that the model
produces useful insights into how non-neutral plasmas and self gravitating systems
behave \cite{antoni1995}.

In this paper, we study the dynamics of an array of $N$ pendula with long-range interacting bobs. By considering long
pendula with flat bobs undergoing small oscillations and having parallel planes of rotation, we produce a model
related to the HMF model through a coordinate transformation. The transformation introduces a 
dependence on the indices of the particle labels. A cartoon of the physical picture is shown in
Fig.~\ref{fig:system}. 
The index dependence in the Hamiltonian, that will be described in detail in the next
section, is a consequence of the pendula pivots being
slightly offset from one another and appears as a phase in the cosine term of the HMF
model. It inspires the investigation of non-equilibrium ``repulsive'' behavior in the angle coordinate frame where
we find an interesting quasistationary state when the 
angles of the pendula are initially ordered according to
their indices. We find the clustered
positions in the usual HMF coordinate frame (biclusters), but in the angle coordinate frame clustering is only
found for the initially ordered angles and, unlike the biclusters, these are clearly quasistationary
states. A quasistationary state is defined as a dynamical state that can persist for a length of time 
which goes to infinity as the thermodynamic limit is approached \cite{campa2009}.
In addition to discussing the clustered angle states exhibited by the system, we also solve for the
canonical partition function in the pendulum angle coordinate frame, finding that in equilibrium with a
heat bath, the probability distributions of the angles can be described by the original HMF model.
This finding is similar to the work done by \cite{campa2000} on a model sometimes called the HMF
$\alpha$-model. In the HMF $\alpha$ model, a $1/r_{ij}^\alpha$ dependence between the classical
spins is introduced \cite{campa2009,campa2003,anteneodo1998,tamarit2000,cirto2014}, where $r_{ij}$ is the
distance between the $i^{\mathrm{th}}$ and $j^{\mathrm{th}}$ spins on a lattice.  Though the
physical motivations behind studying these various models can be very different, it is interesting that their equilibrium behavior is the
same or nearly the same. We believe that the work in this paper further suggests that the HMF
model universally describes an entire class of long-range interacting systems in equilibrium.

\section{The Model}

\subsection{Coordinates}

In Fig.~\ref{fig:system}, we show an array of
pendula rotating in the same plane with bobs that interact through a long-range
potential. If we consider the case where all the pendula only undergo small oscillations, we may
write the horizontal location of the $i^{\mathrm{th}}$ particle, $x_i$, as $x_i = id +
\ell\theta_i$, where $d$ is the distance between the pivots of neighboring 
pendula and $\ell$ is the length
of each pendulum. The small $\theta$ regime makes the problem one dimensional in $x$. We choose
periodic boundary conditions and rescale the system by $2\pi/Nd$ so that 
\begin{equation}
    \label{eqn:scalex}
    x\rightarrow \frac{2\pi}{Nd} x
\end{equation}
making the total system length a
dimensionless $2\pi$ where N is the number of particles in one period. 
We will refer to a periodic space with length $2\pi$ as a unit
circle. The position of the $i^{\mathrm{th}}$ particle (bob) is now 
\begin{equation}
    \label{eqn:scaledxi}
    x_i = \frac{2\pi i}{N} + \frac{2\pi}{N}\frac{\ell}{d}\theta_i
    .
\end{equation}
For reasonable choices of $\ell$ and $d$ ($\ell/d << N$), the second term on the RHS is suitably small such that the
Hamiltonian can be written with terms that are quadratic in $\theta$. However, we are primarily interested in
a regime where $\ell/d \rightarrow \infty$ as the thermodynamic limit is approached. Physically this
corresponds to the small oscillations of very long pendula with suspension points that are close
together compared to their lengths. In order to simplify the calculations that follow, we define
$\phi_i$ to be the last term on the RHS of Eq.~(\ref{eqn:scaledxi}), namely $\phi_i \equiv 2\pi \ell \theta_i / Nd $.
Given the choice of large $\ell/d$, $\phi_i$ can take any value in the
range $[0,2\pi)$. This is only true because $\ell/d$ is large, \textit{not} because the
$\theta_i$s are.  In terms of $\phi_i$, the positions can be rewritten as
\begin{equation}
    \label{eqn:simplescaledxi}
    x_i = \frac{2\pi i}{N} + \phi_i 
    .
\end{equation}

\subsection{Density Approximation}
We have not yet explicitly stated the physical mechanism through which the bobs interact. 
Connecting the interactions with specific physical motivations should 
be discussed with some discretion because the development of the model leaves these motivations up to some
freedom of interpretation. Imagine that the bobs all carry some charge. We will not distinguish between
particles in any other way than their indices, so in the case where all particles carry the same charge,
repulsive behavior is expected. On the other hand one could make the bobs attract one another,
which could be thought of as the self-gravitating case. To be solvable, the model requires some
simplifications. For the sake of brevity we will speak of the particle charge or mass density as the ``density''.

The approximation that we invoke is similar to that used when justifying the HMF model
(Eq.~(\ref{eqn:HMF})) to describe free particles in a
one-dimensional ring \cite{antoni1995,levin2014}. The distribution of the bobs is such the mass
density, $\rho(x)$, is given by
\begin{equation}
    \rho(x) = \sum_{i=1}^{N}\delta(x-x_i) - \frac{1}{2\pi}   
    .
\end{equation}
The constant $1/2\pi$ subtracted from the delta function is necessary to
produce a meaningful expression for the potential $\Phi$ and corresponds to the inclusion of a neutralizing (of opposite sign)
homogeneous background density. Restricting the problem further to that of solving
Poisson's equation for a one-dimensional potential
physically amounts to choosing large and flat bob geometries oriented with their smallest axis 
parallel to the $x$ axis. Writing the delta function as a cosine Fourier series, Poisson's equation becomes
\begin{equation}
    \label{}
    \nabla^2\Phi(x) = \frac{\gamma}{\pi}\sum_{i=1}^{N}\sum_{n=1}^{\infty}
    \cos{[n(x-x_i)]}
    .
\end{equation}
The parameter $\gamma$ contains the particle (bob) charge or mass and becomes the interaction strength in the Hamiltonian.
We can see that the zeroth-order term in the Fourier series canceled the constant neutralizing
background that was superficially added.  

The most important simplification in this paper is truncating the sum of the Fourier coefficients
used to represent the delta function after the $n = 1$ coefficient. 
Antoni et al. defend the truncation by asserting that the ``large scale collective
properties'' do not greatly change when higher order terms of the sum (including interactions at the
smaller length scales) are included, and discuss the
consequences of the approximation in some detail \cite{antoni1995}.
The simplification also warrants a brief discussion of the way that it could be physically 
interpreted. The truncation of the sum is equivalent to smearing out the density of
each particle over the system so that it is peaked at its given location, $x_i$, but also
having a negative density peak on the opposite side of the unit circle. This could be thought
of as doubling the number of particles and enforcing that each particle has a negative partner that
always remains on the opposing side of the unit circle. After this doubling, the now nebulous masses are dispersed such that 
a pair's density is described by a cosine function with the positive
peak centered at $x_i$. 

\subsection{Solving Poisson's Equation}
Integrating Poisson's equation once, we obtain:
\begin{equation}
    \label{}
    \nabla\Phi(x) = \frac{\gamma}{\pi}\sum_{i=1}^{N}
    \left\{
        \sin{(x-x_i)} + c_1
    \right\}
    .
\end{equation}
In order to determine the constant $c_1$ from the integration, the physical picture should
be examined. A sensible requirement is that when all of the bobs are hanging at their equilibrium
positions, directly below their
pivot (all $\phi_i=\theta_i=0$), the net force experienced by any bob is zero. This is a valid
requirement if the bobs are attractive or repulsive, the only difference being that the
configuration would be unstable or stable, respectively.  The force that the
$j^{\mathrm{th}}$ particle experiences when $\phi_j$ and all $\phi_i$ are zero is given by
\begin{equation}
    \label{}
    -\nabla\Phi\left(x_j \right) = -\frac{\gamma}{\pi}\sum_{i=1}^{N}
    \left\{
        \sin
        {
            \left[
                \frac{2\pi (j-i)}{N}
            \right]
        } + c_1
    \right\}
    .
\end{equation}
The sum $\sum_i \sin{[2\pi (j-i)/N]}$ equals zero for any $j$, so $c_1$ must be zero.
Integrating once more to obtain the potential yields 
\begin{equation}
    \label{}
    \Phi(x) = \frac{\gamma}{\pi}\sum_{i=1}^{N}
    \left[
        c_2-\cos{\left(x-\frac{2\pi i}{N} - \phi_i \right)}
    \right] 
    .
\end{equation}
To determine $c_2$ we stipulate that if all
$\phi_i=0$, then $\Phi(0)=0$. Inserting Eq.~(\ref{eqn:scaledxi}) (or
Eq.~(\ref{eqn:simplescaledxi})) for $x_i$ yields
\begin{equation}
    \label{}
    \Phi(0) = \frac{\gamma}{\pi}\sum_{i=1}^{N}
    \left[
        c_2-\cos{\left(\frac{2\pi i}{N}\right)}
    \right]
    .
\end{equation}
The sum over the cosine is zero, therefore $c_2 = 0$ and we can now write the potential energy of the
$j^{\mathrm{th}}$ particle as

\begin{equation}
    \label{}
    \Phi(x_j) = -\frac{\gamma}{\pi}\sum_{i=1}^{N}
    \cos{\left[\frac{2\pi (j-i)}{N}+\phi_j - \phi_i \right]}
    .
\end{equation}

\noindent 
\subsection{The Hamiltonian}
The Hamiltonian can be written as 
\begin{equation}
    \label{}
    H = H_0 + H_I
    ,
\end{equation}
where $H_0$ is the kinetic energy piece 
\begin{equation}
    \label{}
    H_0 = \sum_{i=1}^{N} \frac{p_i^2}{2}
    ,
\end{equation}
and
\begin{equation}
    \label{}
    H_I = -\frac{\gamma}{2N}\sum_{i,j=1}^{N} 
    \cos{
        \left[
            \frac{2\pi (i-j)}{N}
            +
            \phi_i-\phi_j
        \right]
        }
\end{equation}
is the interaction piece, so
\begin{equation}
    \label{eqn:ourH}
    H = \sum_{i=1}^{N} \frac{p_i^2}{2}
    -\frac{\gamma}{2N}\sum_{i,j=1}^{N} 
    \cos{
        \left[
            \frac{2\pi (i-j)}{N}
            +
            \phi_i-\phi_j
        \right]
        }
        .
\end{equation}
The mass of the bobs has been set to unity, $\gamma$ is the interaction
strength, a factor of $1/2$ accounts for the double counting, and the $1/\pi$ coefficient in the
potential energy has been absorbed into $\gamma$. The factor of $1/N$ is a rescaling of the
potential energy that ensures that as the thermodynamic limit is approached, the potential energy of
the system does not diverge. The $1/N$ scaling is known as the Kac prescription 
\cite{teles2012}. 
The Kac serves to keep both the energy and
entropy of a system proportional to the number of particles in the system, an important prerequisite
for phase transitions \cite{campa2009}.

\subsection{Relationship to the HMF model and the Spin Interpretation}
\label{sub:relationship_to_the_hmf_model_and_the_spin_interpretation}

Due to the simple bijective relationship between $x_i$ and $\phi_i$ one can simply solve the
equations of motion for the HMF model and find the dynamics for $\phi_i$ via the coordinate transform
$x_i \rightarrow \phi_i$.  Previously it was mentioned that the HMF model is used to describe free particles
on a ring with long-range repulsion or attraction, as well as describing a classical $XY$ spin model.
The $\theta_i$ played the role of either the position of the $i^{\mathrm{th}}$ particle on the
ring or the orientation of the $i^{\mathrm{th}}$ spin. Therefore, it is interesting to speculate
about the type of spin system the model describes in the $\phi_i$ picture. Thus far, the rescaled
angle $\phi_i = 2\pi \ell \theta_i / Nd$ describes the distance of a pendulum bob from the point
directly below its pivot, but it could also be interpreted as the orientation of spin. In the spin
interpretation of Eq.~(\ref{eqn:ourH}), the potential energy of the $i^{\mathrm{th}}$ and
$j^{\mathrm{th}}$ spin pair depend on both their relative orientation as well as the difference
between their indices. In the following discussion it will sometimes be
convenient to speak about $\phi_i$ in the spin language. 

\begin{figure}[h]
    \centering
    \includegraphics[width=8.0cm]{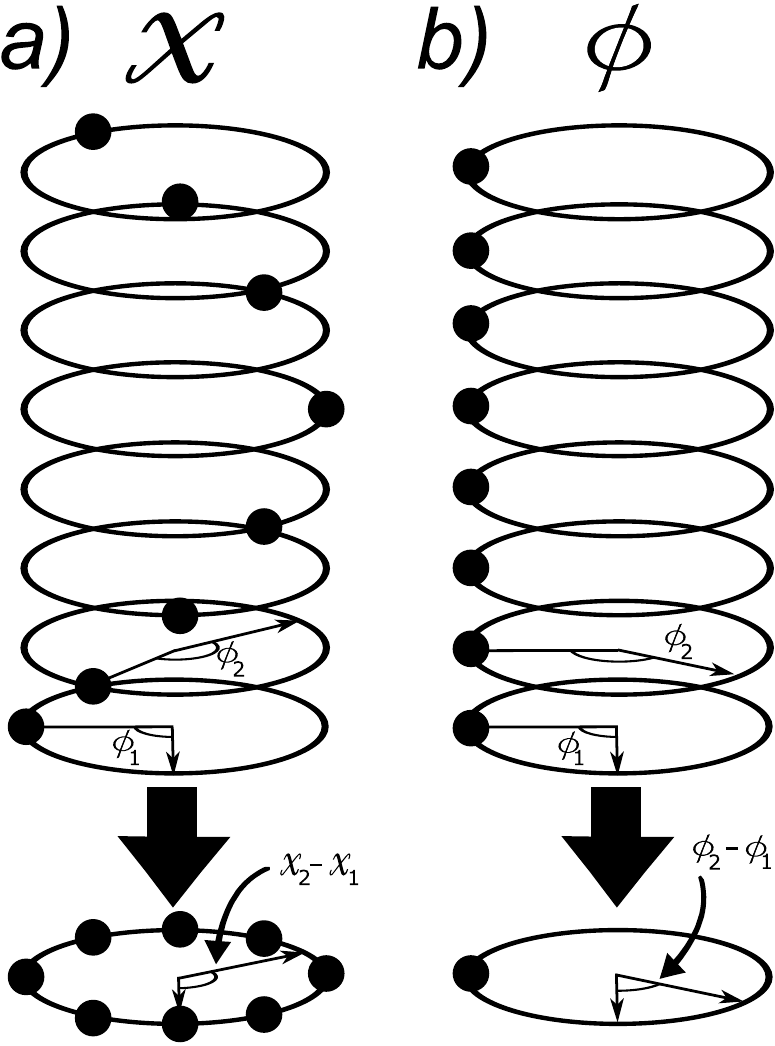}
    \caption{Example of a system of $N=8$ particles when viewed as individual spins in the a) $x$
    coordinate frame, and b) the $\phi$ coordinate frame. 
    a) In the $x$ coordinate frame the direction of the $i^{\mathrm{th}}$ spin given by the angle 
    $x_i$ is expressed as $x_i=2\pi i/N +\phi_i$. Alternatively, $x_i$ could be thought of as the
    position of $i^{\mathrm{th}}$ particle on the unit circle, shown at the bottom of the figure
    as the projection of all positions onto the horizontal plane. 
    The black circles on the rings in the figure
    mark the location of the pendulum pivots at $2\pi i/N$ in $x$. 
    b) Twisting the column of rings in a) such that the pivots are aligned transforms the system into
    the $\phi_i$ coordinate frame. 
    In this picture, the direction of the $i^{\mathrm{th}}$ spin is just given by the angle
    $\phi_i$.
    \label{fig:x_phi_mag}}
\end{figure}

We will prove that in the $\phi$ picture, the system in equilibrium with a heat bath is equivalent to the
HMF model (the $x$ picture) in equilibrium with a heat bath by
solving the partition function in the $\phi_i$ coordinate frame. In the process of simplifying 
the Hamiltonian to solve for the partition function, we will find
expressions of the form $\cos{\phi_i}$ and $\sin{\phi_i}$ which we talk about as the horizontal and
vertical components of a magnetization $\vec{m}_i = (\cos{\phi_i},\sin{\phi_i})$.  It could easily
be stated that in the spin analogy, the $\phi_i$ are orientations of the spins, but we should make a more
concrete connection between this idea and the original presentation of the model. We would like to remind the
reader that even though the angles $\theta_i$ of the pendula are small, the long suspensions of the
bobs ($\ell$) allow $\phi_i$ to cover the entire system which, rescaled, has dimensionless length
$2\pi$. The system is also periodic, so the bobs can be thought of as moving on
a unit circle where the position of the $i^{\mathrm{th}}$ bob is $x_i=2\pi i/N + \phi_i$.  
In order to think of $\phi_i$ as the spin orientations, 
we start by considering each bob as living on its own individual unit circle. An example of these
unit circles is shown in Fig.~\ref{fig:x_phi_mag},
a visual aid to the following.  Imagine stacking horizontal circles in the vertical
direction and rotating each by an angle $2\pi/N$ from the one below. The projection of
these circles onto the horizontal plane would be the system viewed in $x$, i.e. the HMF model.
If we twist the stack so there is no rotation between adjacent circles and then project onto the
horizontal plane it creates the picture viewed in $\phi$, where the pivot points are all aligned.
The reason for this artificial construction of stacked circles is partly to pictorially depict the
transformation between $x$ and $\phi$ and partly to show how $\vec{m}_i$ (as defined) is
just the orientation of the $i^{\mathrm{th}}$ spin in the $\phi$ picture. Said
differently, each circle in the $\phi$ picture represents a spin with an orientation in the
horizontal plane determined by $\phi_i$; an infinite-range classical
mean field spin model described by Eq.~(\ref{eqn:ourH}).

\section{Equilibrium}

In this section, we solve for the canonical partition function, in the $\phi$ coordinate frame using
the Hamiltonian in Eq.~(\ref{eqn:ourH}) and show that, in equilibrium, the HMF model describes the
angles of long pendula with long-range interacting bobs. In order to solve the configurational piece
of the partition function the Hamiltonian must be modified. Using the cosine and sine sum and
difference identities twice, the potential interaction piece of the Hamiltonian $H_I$ can be written
as

\begin{multline}
%    \frac{-\gamma}{2} - \frac{\gamma}{2N} \sum_{i,j} 
    H_i =
      \frac{-\gamma}{2N} \sum_{i,j} 
        \Bigg\{
            \cos{\left[ \frac{2\pi (i-j)}{N} \right]}
            [
                \cos{\phi_i}\cos{\phi_j} 
            \\
                +
                \sin{\phi_i}\sin{\phi_j}
            ]
            \\
             -
            \sin{\left[ \frac{2\pi (i-j)}{N} \right]}
            \left[
                \sin{\phi_i}\cos{\phi_j} 
                -
                \cos{\phi_i}\sin{\phi_j}
            \right]
        \Bigg\}
        .
\end{multline}

\noindent
The coefficients in the Hamiltonian $\cos{[2\pi (i-j)/N]}$ and $\sin{[2\pi (i-j)/N]}$ should be
thought of as matrices with components $C_{ij}$ and $S_{ij}$ respectively. The Hamiltonian can
now be written in the form

% single line
%\begin{equation}
%    \label{}
%    \frac{\gamma}{2N}\sum_{i,j}
%    \left(
%        \cos{\phi_i}C_{ij} \cos{\phi_j} + \sin{\phi_i}C_{ij}\sin{\phi_j} - \sin{\phi_i}S_{ij}\cos{\phi_j} +
%        \cos{\phi_i}S_{ij}\sin{\phi_j}
%    \right),
%\end{equation}
\begin{multline}
    \label{}
    \frac{\gamma}{2N}\sum_{i,j}
    (
        \cos{\phi_i}C_{ij} \cos{\phi_j} + \sin{\phi_i}C_{ij}\sin{\phi_j} 
        \\
        - \sin{\phi_i}S_{ij}\cos{\phi_j} + \cos{\phi_i}S_{ij}\sin{\phi_j}
    ),
\end{multline}

\noindent
which is suggestive because it can be regarded as the matrix equation

\begin{multline}
    \label{}
    H_I = \frac{\gamma}{2N}
    \Bigg[
        (\cos{\phi_1},\cos{\phi_2},...,\cos{\phi_N}) 
        C 
        \begin{pmatrix} 
            \cos{\phi_1} \\ 
            \cos{\phi_2} \\ 
            \vdots \\ 
            \cos{\phi_N} 
        \end{pmatrix} 
        \\
        + 
        (\sin{\phi_1},\sin{\phi_2},...,\sin{\phi_N}) 
        C 
        \begin{pmatrix} 
            \sin{\phi_1} \\ 
            \sin{\phi_2} \\ 
            \vdots \\ 
            \sin{\phi_N} 
        \end{pmatrix} 
        \\
        - 
        (\sin{\phi_1},\sin{\phi_2},...,\sin{\phi_N}) 
        S 
        \begin{pmatrix} 
            \cos{\phi_1} \\ 
            \cos{\phi_2} \\ 
            \vdots \\ 
            \cos{\phi_N} 
        \end{pmatrix} 
        \\
        + 
        (\cos{\phi_1},\cos{\phi_2},...,\cos{\phi_N}) 
        S 
        \begin{pmatrix} 
            \sin{\phi_1} \\ 
            \sin{\phi_2} \\ 
            \vdots \\ 
            \sin{\phi_N} 
        \end{pmatrix} 
    \Bigg]
    .
\end{multline}

\noindent
It is helpful to consider the particles positions on the
unit circle with respect to their pivot ($\phi$) as magnetizations. 
Defining

\begin{equation}
    \label{}
    \vec{m}_i \equiv (\cos{\phi_i},\sin{\phi_i}),
\end{equation}

\noindent 
and with $m^T_{\mu} = (m_{0,\mu},m_{1,\mu},...,m_{N-1,\mu})$ where $\mu$ holds the place of $x$ or
$y$, the Hamiltonian becomes
\begin{multline}
    \label{}
    H_I = \frac{\gamma}{2N}
    (
        m^T_x C m_x + m^T_y C m_y 
        \\
        - m^T_y S m_x + m^T_x S m_y
    )
    .
\end{multline}

A closer examination of the structure of the coefficient matrices $C$ and $S$ indicates that they take
the special form of circulant matrices, and thus can be simultaneously diagonalized by a unitary
matrix $U$.  A circulant matrix has the form 
\begin{equation}
    \label{eqn:circmatrix}
    \left(
    \begin{matrix}
        a_1     & a_2    & a_3    & \ldots & a_{N-1} & a_N     \\
        a_N     & a_1    & a_2    & \ldots & a_{N-2} & a_{N-1} \\
        a_{N-1} & a_N    & a_1    & \ldots & a_{N-3} & a_{N-2} \\
        \vdots  & \vdots & \vdots & \ddots & \vdots  & \vdots  \\
        a_3     & a_4    & a_5    & \ldots & a_1     & a_2     \\
        a_2     & a_3    & a_4    & \ldots & a_N     & a_1     \\
    \end{matrix}
    \right)
    ,
\end{equation}
a special kind of Toeplitz matrix, where each subsequent row is a cyclic
permutation of the row above or below it. Any matrix $A$ with elements $a_{ij}$ that can be written
in terms of some function $f(i-j)$ is a circulant matrix. 
Because a circulant matrix is a normal matrix it can be diagonalized by a unitary matrix. 
We show that $C$ and $S$ are simultaneously diagonalizable by showing that they commute, i.e.  
$[C,S]=0$ where $[C,S] = CS-SC$. Starting with the second term, $-SC = S^TC^T=(CS)^T$ which is found
by arguing that $C$ is symmetric since cosine is an even function and does not change under the exchange
of $i$ and $j$, whereas $S$ is odd because sine is
an odd function and does change sign under exchange of $i$ and $j$. The
commutation becomes $[C,S] = CS + (CS)^T$. Also, an odd function multiplied by an even
function results in an odd function so the entire matrix $CS$ is odd. Therefore
$(CS)^T = -CS$ bringing us to the final expression $[C,S] = CS-CS=0$. We have shown that $C$ and $S$ can be
simultaneously diagonalized by $U$. The matrix $U$ is known for circulant matrices and is called the
Fourier Matrix. 

The matrices $C$ and $S$ can be rewritten as
$C= U^\dagger D^C U$ and $S = U^\dagger D^S U$, where $D^C$ and $D^S$ are diagonal matrices
with diagonal elements that are the eigenvalues of $C$ and $S$, respectively, which we denote as
$\lambda^C_i$ and $\lambda^S_i$. From here on we label the indices $i$ from $0$ to $N-1$. It is worth pointing
out that due to $S$ being antisymmetric, $U$ must be complex. Equation~(\ref{eqn:matrixCS})
becomes
\begin{multline}
    \label{eqn:matrixCS}
    H_I = \frac{\gamma}{2N}
    \Big(
        m^T_x U^\dagger D^C U m_x + m^T_y U^\dagger D^C U m_y 
        \\
        - m^T_y U^\dagger D^S U m_x + m^T_x U^\dagger D^S U m_y
    \Big)
    .
\end{multline}
We will move back to the index notation using the following relations:
\begin{equation}
    \label{}
    D^{C,S} = \lambda_i^{C,S}\delta_{ij}
    ,
\end{equation}
\begin{equation}
    \label{}
    X_j \equiv \sum_{k=1}^N U_{jk} m_k^x
    ,
\end{equation}
and
\begin{equation}
    \label{}
    Y_j \equiv \sum_{k=1}^N U_{jk} m_k^y.
    ,
\end{equation}
where $X$ is not to be confused with $x$. 
Using the Kronecker delta, we set all $i=j$ since these are the only nonzero terms. 
The Hamiltonian is now given by
\begin{multline}
    \label{}
    H_i = 
    \frac{-\gamma}{2N}\sum_{j=0}^{N-1}
    \Big(
        \|X_j\|^2 \lambda_j^C + \|Y_j\|^2 \lambda_j^C 
        \\
        -Y_j^* X_j \lambda_j^S + X_j^* Y_j \lambda_j^S
    \Big)
    ,
\end{multline}
with $\|X\|= XX^*$ and $\|Y\|= YY^*$. The inclusion of the eigenvalues $\lambda^C$ and
$\lambda^S$ simplifies the Hamiltonian further. We will now solve for $\lambda^C$ and $\lambda^S$.
Looking at the form of a circulant matrix shown in Eq.~(\ref{eqn:circmatrix}) reminds us that the
elements of a circulant matrix can be defined with a single label. We write the single labeled
elements of the cosine and sine matrices respectively as
\begin{equation}
    \label{}
    c_l = \cos{\frac{2\pi l}{N}}
    ,
\end{equation}
and
\begin{equation}
    \label{}
    s_l = \sin{\frac{2\pi l}{N}}
    ,
\end{equation}
where $l=0,1,2,...,N-1$. The eigenvalues, $\lambda^A$, of a $N\times N$ 
circulant matrix $A$ can be written in terms of the
single label elements $a_l$. The $j^{\mathrm{th}}$ eigenvalue of $A$ is known to be 
$\lambda^A_j=\sum_{l=0}^{N-1} a_l \exp{(2\pi i l j / N)}$, where $i$ is $\sqrt{-1}$ (not an index)
and $l=0,1,2,...,N-1$.  Therefore,
\begin{equation}
    \label{}
    \lambda_j^C = \sum_{l=0}^{N-1} \cos{\left(\frac{2\pi l}{N}\right)} e^{2\pi i l j /N}
    ,
\end{equation}
and
\begin{equation}
    \label{}
    \lambda_j^S = \sum_{l=0}^{N-1} \sin{\left(\frac{2\pi l}{N}\right)} e^{2\pi i l j /N}
    .
\end{equation}
Writing cosine and sine in their exponential forms gives
\begin{equation}
    \label{}
    \lambda_j^C = \frac{1}{2} \sum_{l=0}^{N-1} 
    \left[
        e^{i 2 \pi l (j+1)/N}
        +
        e^{i 2 \pi l (j-1)/N}
    \right]
    ,
\end{equation}
\begin{equation}
    \label{}
    \lambda_j^S = \frac{-i}{2} \sum_{l=0}^{N-1} 
    \left[
        e^{i 2 \pi l (j+1)/N}
        +
        e^{i 2 \pi l (j-1)/N}
    \right]
    ,
\end{equation}
The above representations of the eigenvalues show that $C$ and $S$ each have only two non-zero
eigenvalues corresponding to $j=1,N-1$ given by
$\lambda_1^C = \lambda_{N-1}^C = N/2$ and $\lambda_1^S=(\lambda_{N-1}^{S})^* = iN/2$.
The Hamiltonian simplifies greatly to

\begin{equation}
    \label{eqn:simpleH}
    H_I = \frac{\gamma}{2}
        -
        \left(
            \| X_1+iY_1 \|^2 + \| X_{N-1} - iY_{N-1} \|^2
        \right).
\end{equation}

The representation of $H_I$ in Eq.~(\ref{eqn:simpleH}) must be further modified before the partition
function can be found. We do this by splitting the Fourier matrix $U$ into its real
and imaginary components, $a_{ik}$ and $b_{ik}$, given by
\begin{equation}
    \label{}
    a_{ik} \equiv \frac{1}{\sqrt{N}} \cos{\left(\frac{2 \pi i k}{ N} \right)},
\end{equation}
and
\begin{equation}
    \label{}
    b_{ik} \equiv \frac{1}{\sqrt{N}} \sin{\left(\frac{2 \pi i k }{ N} \right)}
    .
\end{equation}
This was done to write the absolute squares in Eq.~(\ref{eqn:simpleH}) in terms of the squares of 
$a_{ik}$ and $b_{ik}$. By noticing that $a_{1k} = a_{(N-1)k}$ and $b_{1k} = -b_{(N-1)k}$ 
we write the configurational partition function as

\begin{multline}
    \label{eqn:partition_pre_hubbard}
    Z_I = A \int d^N\phi
    e^
    {
        \frac{\beta \gamma}{2} 
        \left(
            \sum_k 
            \left[ 
                a_{1k} m_k^x - b_{1k} m_k^y 
            \right]
        \right)^2
    } 
    \\
    \times
    e^
    {
        \frac{\beta \gamma}{2}
        \left(
        \sum_k
            \left[
                b_{1k} m_k^x + a_{1k} m_k^y 
            \right]
        \right)^2
    }
    ,
\end{multline}

\noindent
where $\beta=1/k_B T$. 

The Hubbard-Stratonovich transformation is now applied twice, once to each quadratic quantity in
the partition function. The integration variables introduced through this transformation are $z_1$
and $z_2$ with subscripts for first and second quadratic quantities, respectively. After 
after switching the order of integration, we find
\begin{multline}
    \label{}
    Z_I = 
    \frac{A }{2 \pi \beta \gamma} 
    \int_{-\infty}^{\infty} dz_1 dz_2
    e^
    {
        -(z^2_1+z^2_2)/ 2 \beta \gamma
    } 
    \prod_k
    \\
    \times
    \int_{-\pi}^{\pi} d \phi_k
    e^
    {
        (z_{1}a_{1k}+z_{2}b_{1k})\cos{\phi_k} + (z_{2}a_{1k}-z_{1}b_{1k})\sin{\phi_k}
    }.
\end{multline}
The integration can be performed using the identity 
\begin{equation}
    \label{}
    \int_{-\pi}^{\pi} d\phi e^{\xi \cos{\phi} + \eta \sin{\phi}} = 2\pi I_0 \left( \sqrt{\xi^2 + \eta^2} \right)
\end{equation}
where
\begin{equation}
    \label{}
    \xi^2 + \eta^2 = (z_1 a_{1k} + z_2 b_{1k})^2 + (z_2 a_{1k} - z_1 b_{1k})^2
\end{equation}
which simplifies when $a$ and $b$ are included to
\begin{multline}
    \label{}
    \left[
        z_1 \frac{1}{\sqrt{N}}\cos{\left(\frac{2\pi k}{N}\right)}
        +
        z_2 \frac{1}{\sqrt{N}}\sin{\left(\frac{2\pi k}{N}\right)}
    \right]^2
    \\
    +
    \left[
        z_2 \frac{1}{\sqrt{N}}\cos{\left(\frac{2\pi k}{N}\right)}
        -
        z_1 \frac{1}{\sqrt{N}}\sin{\left(\frac{2\pi k}{N}\right)}
    \right]^2
    \\
    =
    \frac{1}{N}(z_1^2 + z_2^2)
\end{multline}
It is convenient to make a change to polar coordinates by introducing $z=\sqrt{z_1^2 +z_2^2}$, following 
which the partition function can be written as
\begin{equation}
    \label{eqn:adZ}
    Z_I = 
    \frac{A }{ \beta \gamma} 
    \int_{-\infty}^{\infty} dz
    e^
    {
        -z/ 2 \beta \gamma
    } 
    \prod_k
    2 \pi I_0 
        \left(
            \frac{\sqrt{z}}{\sqrt{N}}
        \right)
    .
\end{equation}
Equation~(\ref{eqn:adZ}) is recognized to be an intermediate step
of the solution to the canonical partition function for the HMF model. 
From here we jump to the main results, the details of which are included in the HMF literature
\cite{antoni1995,levin2014,campa2009} . 

\begin{figure}[h]
    \centering
    \includegraphics[width=8.0cm]{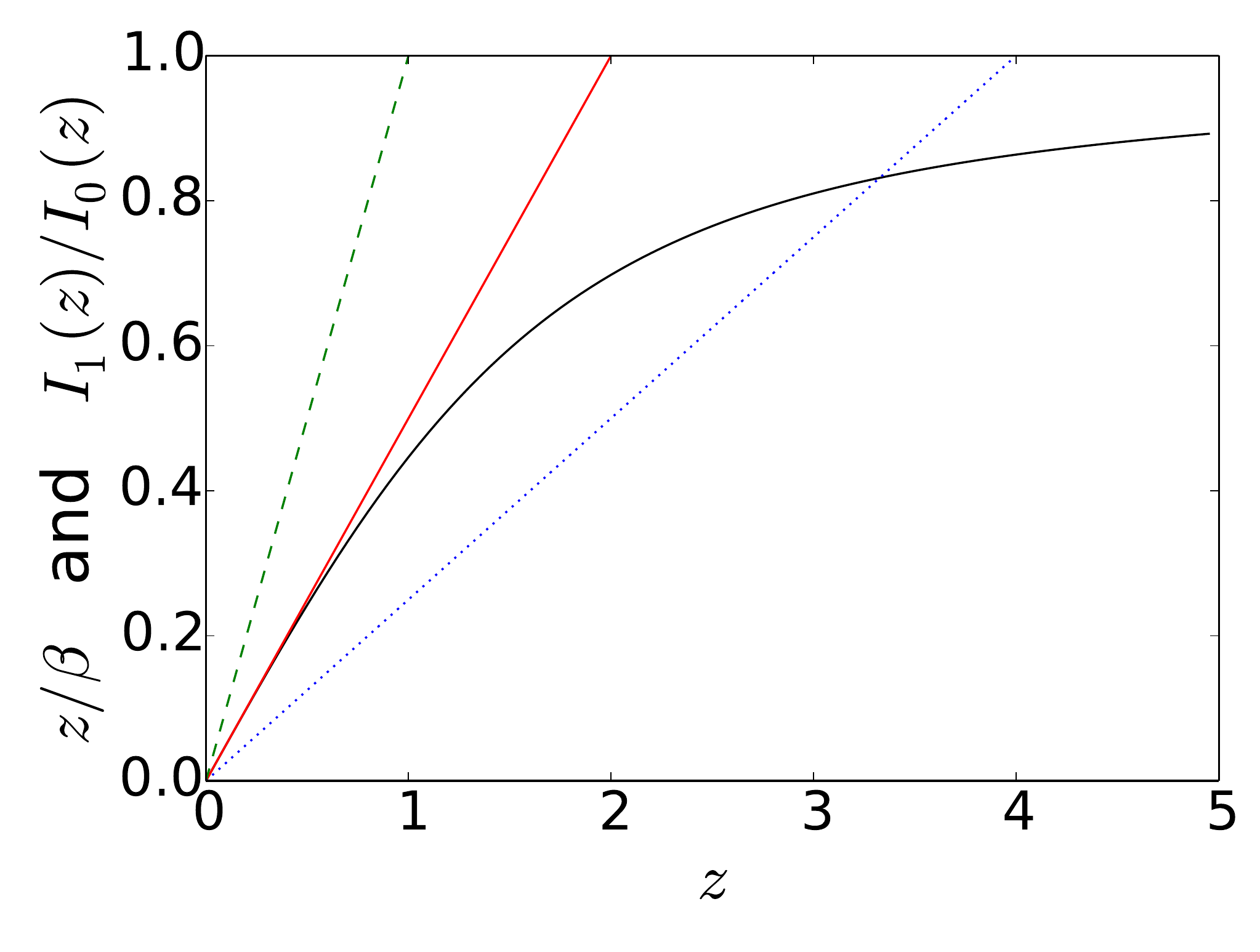}
    \caption{The solid (black) curve is the fraction of modified Bessel functions $I_1(z)/I_0(z)$,
    dashed (green) is $z/\beta$ for $\beta=1$, solid (red) line is $z/\beta$ for $beta=2$,
    dotted (blue) is $z/\beta$ for $\beta=4$, all as a function of $z$.
    (blue) line is $z/\beta$ for $\beta=4$. The values $\beta=1,2,4$ correspond to
    the pre-phase transition, critical, and post-phase transition values in that order.
}
    \label{fig:graphical_z_bessel}
\end{figure}

The integration over $z$ in Eq.~(\ref{eqn:adZ}) can be preformed using the saddle point
approximation. The rescaled free energy per particle follows as
\begin{equation}
    \label{eqn:sup}
    -\beta F = 
    % don't have this because there is no more constant in sum in hamiltonian_I \frac{1}{2}\ln{2\pi/\beta} 
    - \frac{\beta}{2} 
    +
    \inf_z
    \left[
        \frac{- z^2}{2\beta} + \ln{2\pi I_0(z)}
    \right]
\end{equation}
The expression above permits a convenient path to finding the phase transition. Solving for the
minimum values of $z$ in order to satisfy the last term in Eq.~(\ref{eqn:sup}) results in the equation
\begin{equation}
    \label{}
    \frac{z}{\beta} - \frac{I_1(z)}{I_0(z)} = 0
    ,
\end{equation}

\noindent
which can be solved self consistently for $z$ and represented graphically for different values of
$\beta$ as in Fig.~\ref{fig:graphical_z_bessel}.
The reader will see that
after $\beta$ is increased passed the critical value ($\beta=2$) there are two well-defined
solutions.

The Hubbard-Stratonovich transformation decouples
spin-spin (squared terms in the Hamiltonian) contributions
to the partition function at the price of needing to create a linear interaction between each spin with
an auxiliary field $z$ 
\cite{altland2006}.  
Again, a more detailed procedure can
be found in 
\cite{antoni1995,levin2014,campa2009}
where discussion of the internal energy in the
equilibrium state is followed by non-equilibrium behavior of the system prepared in microcanonical
ensembles. Here we will simply touch on the most important point of the equilibrium behavior, being
that for $\beta<2$ the system is paramagnetic but for $\beta \ge 2$ a pitchfork bifurcation
occurs resulting in two stable solutions. At this point there is a discontinuity in the free energy,
a second order phase transition occurs and the system can maintain finite magnetization. In this
case, the order parameter is the total magnetization $\vec{M}=\frac{1}{N}\sum_{j=1}^{N} \vec{m}_i$
where $\vec{m}_i$ was defined to be $(\cos{\phi_i},\sin{\phi_i})$. 

Showing that the canonical partition function in the $\phi$ coordinate frame model 
and the HMF model are equivalent
necessitates a more detailed discussion of the equilibrum behaviors in the $\phi$ frame. Campa et al. 
\cite{campa2009},
in their review of the HMF model, rigorously show ensemble equivalence between the canonical and
microcanoical ensemble of the HMF model. In light of this fact, a large
$N$ microcanonical simulation should be able to produce equilibrium behavior like the phase
transition mentioned above. The temperature in a numerical simulation of a system with many
particles would be ``set'' through a choice
of the initial momenta distribution. In this type of simulation, it is common practice to
compute the order parameter and free energy \cite{dynam_therm_intro,miritello2009,antoni1995},
begging the question: does a large microcaonical simulation of Eq.~(\ref{eqn:ourH}) approximate the
expected equilibrium behavior? Also, since the index-dependent model in equilibrium with a heat bath
can be described by the HMF model, would the dynamics of such a simulation qualitatively
resemble those in the HMF model? The answer to both of these questions is \textit{no} if one were to
find the equations of motions in $\phi_i$ for some large $N$ and then compare them to an HMF model
or the $x$ coordinate frame.  As stated, this discrepancy may appear to detract from our result.
Indeed, it uncovers a conceptual omission in the model, but it is one whose rectification gives
insight into the models ensemble equivalence property of the model, or lack thereof. The omission was in the arbitrary
scaling of $x$ which we will now rectify.  

We introduce the parameter $L$ which generalizes the scaling
in Eq.~(\ref{eqn:scalex}) to
\begin{equation}
    \label{eqn:newscalex}
    x\rightarrow \frac{2\pi L}{Nd} x
    ,
\end{equation}
making the position of the $i^{\mathrm{th}}$ particle
\begin{equation}
    \label{eqn:newscaledxi}
    x_i = \frac{2\pi L i}{N} + \frac{2\pi L}{N}\frac{\ell}{d}\theta_i
    .
\end{equation}
and changing the definition of $\phi_i$ to $\phi_i\equiv 2\pi L\ell\theta_i/Nd$.  It can be shown
that the introduction of $L$ only changes the final result of the partition function by a constant
factor of $L$ due to the enlarged limits of integration.  Numerically, we find is that 
if $L>>1$, then the simulations in $\phi$ closely reproduce the dynamics of
HMF model simulations (dynamics in $x$). Therefore, for large $L$ the mirocanonical simulations can
approximate equilibrium and the
answers to the previous questions - does a large microcaonical simulation of Eq.~(\ref{eqn:ourH}) approximate the
expected equilibrium behavior, and since the index-dependent model in equilibrium with a heat bath
can be described by the HMF model, would the dynamics of such a simulation qualitatively
resemble those in the HMF model? - becomes \textit{yes}. 
Alternatively, the coordinate frame inequivelence is most extreme for small $L$. 
These numerical results were found using initial conditions that are randomly distributed $\phi_i$
about the domain $[-L\pi,L\pi)$.  It should be stated that for the rest of this paper we work with
$L=1$ becuase we are inetersed in cases where the $\phi$ coordinate frame is markedly differnet than
the $x$ coordinate frame. 

%%%%%%%%%%%%%%%%%%%%%%%%%%%%%%%%%%%%%%%%%%%%%%%%%%%%%%%%%%%%%%%%%%%%%%%%%%%%%%%%%%%%%%%%%%%%%%%%%%%%%%%%%%

\section{Non-Equilibrium Results}

For a system of pendula, it is interesting to study an initial configuration where all pendula are 
set to random small displacements from $\phi_i=0$. Specifically we initialize the $i^{\mathrm{th}}$
pendulum angle, $\phi_i$, randomly in the range $[-\pi/N,\pi/N)$. In $x$ the indices
are ordered in $x$ such that $x_1<x_2<x_3<...<x_N$ and the $i^{\mathrm{th}}$ bob is randomly
distributed in the range $[2\pi i/N-\pi/N,2\pi i/N+\pi/N)$. It is possible to make some general
statements about the dynamics of this configuration in $x$ using the equations of motion. 
Expressing the Hamiltonian with terms that are quadratic in $\phi$ yields

\begin{multline}
    \label{}
    H_I = \frac{\gamma}{2N}\sum_{ij} 
    \Bigg\{ 
        \sin
        {
            \left[
                \frac{2 \pi (i-j)}{N}
            \right]
        }
            (\phi_i - \phi_j)
        \\
        -\cos
        {
            \left[
                \frac{2 \pi (i-j)}{N}
            \right]
        }
        \left(
            1-\frac{\phi_j^2}{2} - \frac{\phi_i^2}{2}+\phi_i \phi_j
        \right)
    \Bigg\}
    .
\end{multline}

\noindent
With this expression, the equations of motion for the $i^{\mathrm{th}}$ particle can be written as 

\begin{equation}
    \label{eqn:eqn_motion}
    \ddot{\phi}_i = \dot{p}_i = \frac{\partial H}{\partial \phi_i},
\end{equation}

\noindent
from which we obtain

\begin{multline}
    \label{}
    \ddot{\phi}_i = \frac{-\gamma}{2N}\sum_{j} 
    \Bigg\{
        \cos{\left[\frac{2\pi(i-j)}{N}\right]}(\phi_j-\phi_i) 
        \\
        +
        \sin{\left[ \frac{2\pi (i-j)}{N} \right]}
    \Bigg\}
        .
\end{multline}

\noindent
In the above equation, the last term and the $\cos{[2\pi(i-j)/N]}\phi_i$ term sum to zero,
leading to

\begin{equation}
    \label{}
    \ddot{\phi}_i = \frac{-\gamma}{2N} \sum_j  \cos{\left[\frac{2\pi(i-j)}{N}\right]}\phi_j
    .
\end{equation}

\noindent
Using the difference formula, we write the acceleration as

\begin{equation}
    \label{eqn:initF}
    \ddot{\phi}_i = \frac{-\gamma}{2} 
    \left[
        \cos{\left( \frac{2\pi i}{N} \right)} \langle \mu_1 \rangle
        +
        \sin{\left( \frac{2\pi i}{N} \right)} \langle \mu_2 \rangle
    \right]
    ,
\end{equation}

\noindent
where $\mu_1 = \phi_j\cos{(2\pi i/N)}$ and $\mu_2 = \phi_j\sin{(2\pi i/N)}$.  The mass (moment of
inertia) has been set to unity so the above expression is the force as a function of index,
$\ddot{\phi_i} = F(i)$. 
If $\langle \mu_1
\rangle$ and $\langle \mu_2 \rangle$ are known, then the initial dynamics of the system are
elucidated by Eq.~(\ref{eqn:initF}), but in the case of randomly initialized $\phi_i$ the 
$\langle \mu_1 \rangle$ and $\langle \mu_2 \rangle$ are also random and can be different from one another in
both magnitude and sign. However, a general description of the results can be given without exactly knowing these
coefficients. Equation~(\ref{eqn:initF}) shows that the initial force on a given particle
depends on its position because the indices are ordered in $x$. In the continuum (thermodynamic
limit), the force takes the form 

\begin{equation}
    \label{eqn:initFx}
    F(x) \equiv \frac{-\gamma}{2} 
    \left(
        \langle \mu_1 \rangle \cos{x} + \langle \mu_2 \rangle \sin{x}
    \right)
    .
\end{equation}

\begin{figure}[h]
    \centering
    \includegraphics[width=8.0 cm]{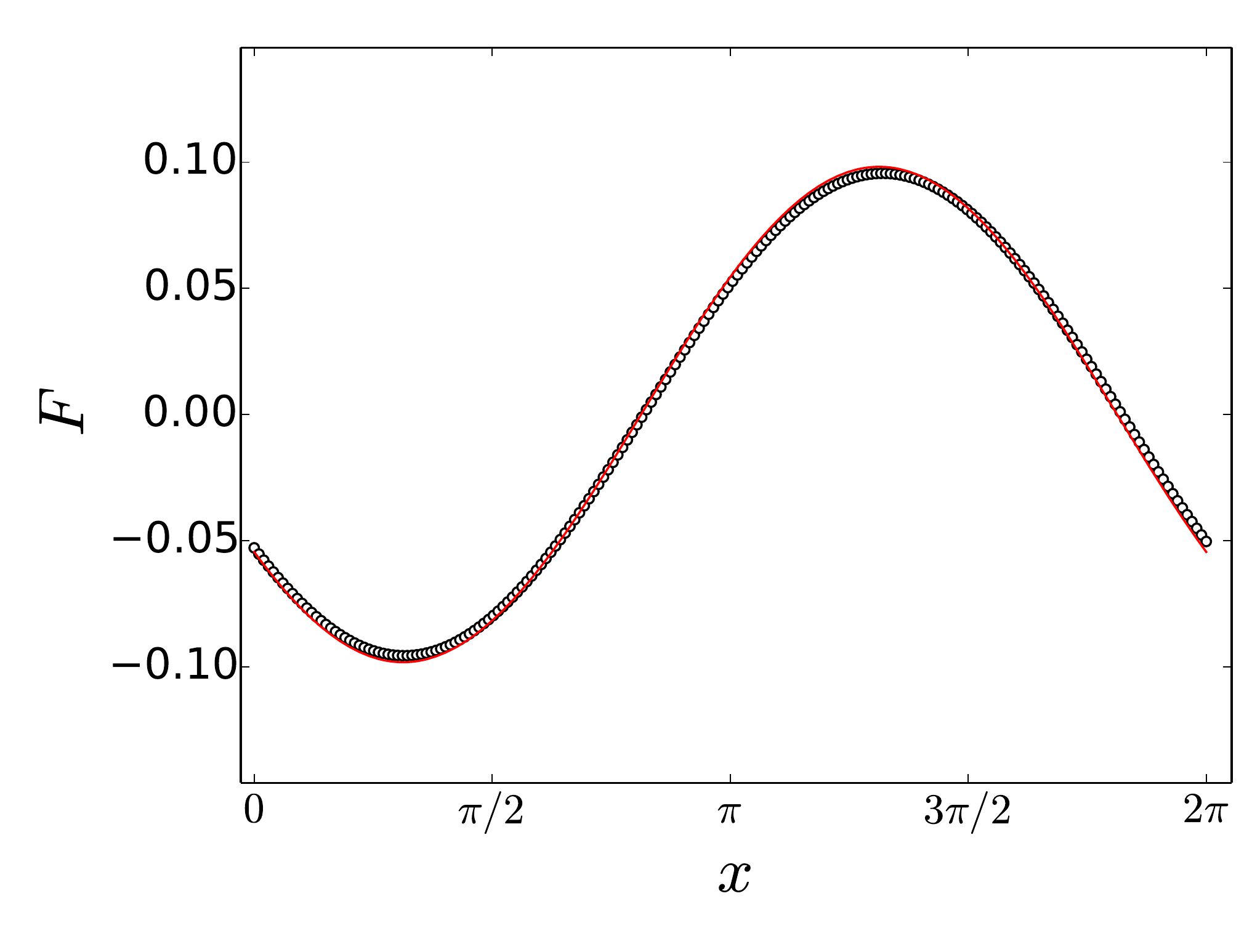}
    \caption{Numerical (blue) and theoretical (red) value of the $t=0$ force 
        felt by each particle as
        a function of its position. This configuration was made with $N=200$ and $\gamma=10$. 
        The initial $\phi_i$ used for the
        numerical calculation were chosen randomly in the range $[0,2\pi/N)$ which was restricted to 
        positive values so that the sign
        of the initial $\langle \mu_1 \rangle$ and $\langle \mu_2 \rangle$ were known to be
        positive. This is not significantly different than
        when the range of $\phi_i$ is centred about
        0. The
        theoretical curve is fitted using Eq.~(\ref{eqn:initFx}) with $\langle \mu_1 \rangle =
        0.0109$ and $\langle \mu_2 \rangle = 0.0163$. \label{fig:initF}}
\end{figure}

\noindent
Therefore $\langle \mu_1 \rangle$ and $\langle \mu_2 \rangle$ partly play the role of the amplitude of
this force as a function of $x$, but also can shift the $\cos{x}+\sin{x}$ spatial dependence, which is periodic
over the system length. In Fig.~\ref{fig:initF}, 
we show a fit of the force as a function of $x$ using Eq.~(\ref{eqn:initFx}) as well as the actual
force calculated for an example set of initial conditions. The domain in Fig.~\ref{fig:initF} can be 
split into two pieces (independent of $\mu$)- one where the particles
experience a positive force, the other in which the particles experience a negative force. As time
is increased, the movements of the particles evolve the coefficients $\langle \mu_1 \rangle$ and
$\langle \mu_2 \rangle$ in such a way that the magnitude of the force decreases to zero for all
particles and then switches sign complementary to the original force. This results in a standing
compression wave of the particles with a wavelength $2\pi$. The compression wave is not stable and
eventually two clusters form about each node. These two clusters are often referred to as
a ``bicluster'', or the antiferromagnetic state in the HMF model, and have been explained 
by Barr\'{e} et al. by analysing the Vlasov equation. They find that the
initial compression wave (referred to by a different name) 
creates an effective double-well potential giving rise to the bicluster
\cite{barre2001}. The question of the bicluster stability has not yet been definitely answered, but for a
detailed discussion we refer the reader to Leyvraz et al. \cite{leyvraz2002}.
Given the simple mapping between the $\phi$ and $x$ coordinate frames, 
we should also be able to show the initial form of the force in $x$ as well.
As presented in Eq.~(\ref{eqn:ourH}), the Hamiltonian in the $x$ coordinate frame only differs from
the HMF model by a constant $\gamma/2$. In $x$, $H_I$ is
\begin{equation}
    \label{}
    H_I= \frac{\gamma}{2N}\sum_{i,j=1}^{N}\cos{( x_i - x_j )}.
\end{equation}
Using the difference identity, we find the equations of motion for the $i^{\mathrm{th}}$ particle 
to be
\begin{equation}
    \label{}
    \ddot{x}_i
    \frac{-\gamma}{2N}
    \left(
        -\sin{x_i}\sum_j \cos{x_j} + \cos{x_i}\sum_j \sin{x_j}
    \right).
\end{equation}
The sums over cosine and sine of $x_j$ play the same role as $\langle \mu_1 \rangle$ and 
$\langle \mu_2 \rangle$, and the force at a given position $x_i$ is clearly of the same form as that shown in
Eq.~(\ref{eqn:initF}).
%\begin{equation}
%    \label{eqn:eqn_motion}
%    \ddot{x}_i = \dot{p}_i = \frac{\partial H}{\partial \phi_i},
%\end{equation}

Depending on $\langle \mu_1 \rangle$ and $\langle \mu_2 \rangle$, all $\phi_i$ oscillate about zero with amplitudes and
phases that depend on their location $x_i$ as discussed above. As the clustering in $x$ begins, the
$\phi_i$ begin to spread out over the full domain $[0,2\pi)$ and continue to do so until it is covered. The more
interesting case in $\phi$ is when all $\phi_i$ are initially randomly distributed in ranges that depend
on their index, specifically when $\phi_i$ are chosen in the ranges. $[2\pi i/N-\pi/N,2\pi
i/N+\pi/N)$ so that $\phi_1<\phi_{2}<\phi_{3}<...<\phi_{N}$. It should be noted that in this new
configuration the
dynamics in $x$ are nearly identical to the configuration previously discussed for
ordered $x_i$. The dynamics in $\phi$ differ \textit{drastically} between the two cases
though. In this ordered angle case, we find some interesting grouping of the scaled angles. 

\begin{figure}[h]
    \centering
    \includegraphics[width=8.0 cm]{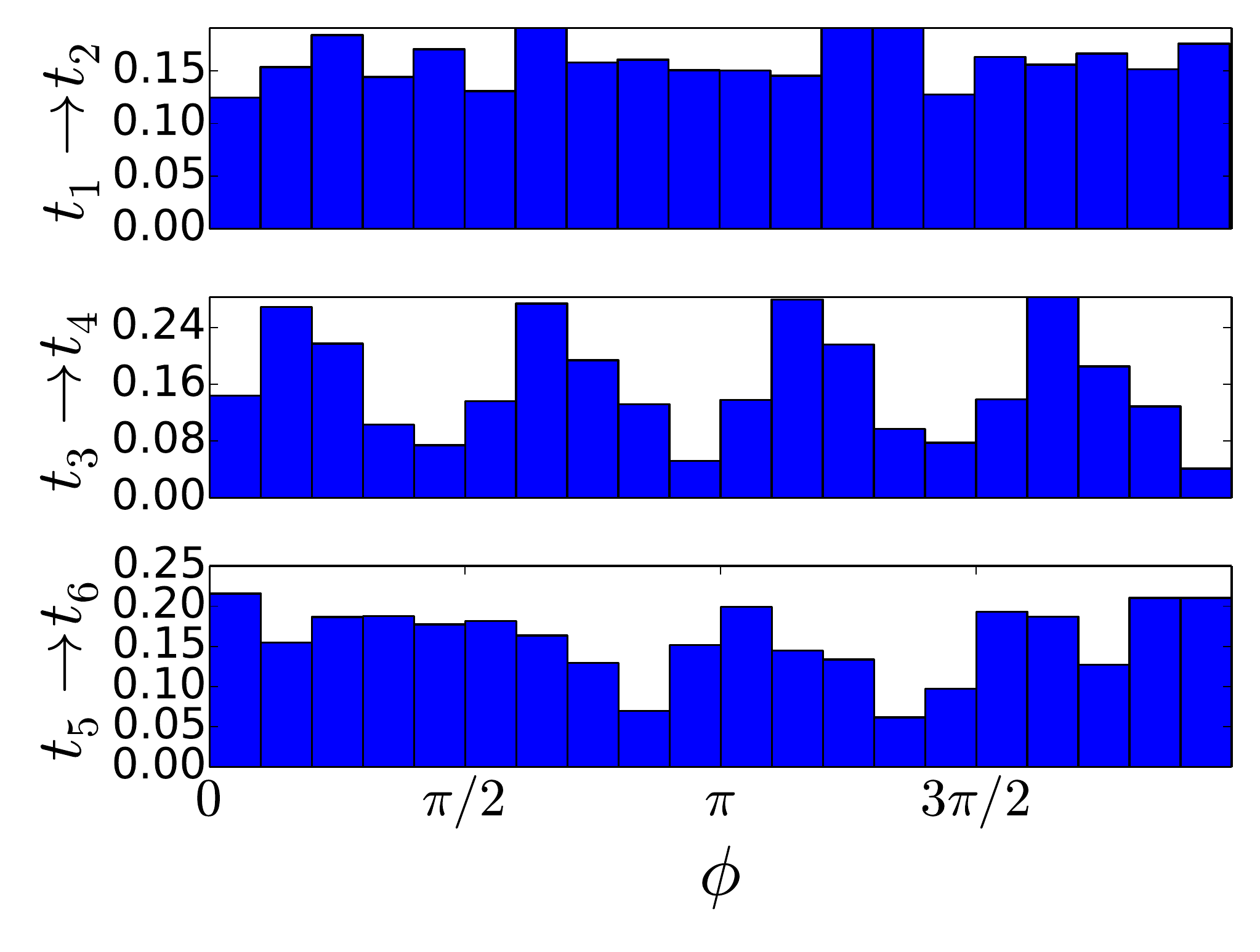}
    \caption{These three histograms are made by binning $\phi_i$ of 50 particles 
    over three different periods of time with $\gamma=10$ . Going from top to bottom each period of time
    belongs to the dynamical regime of: standing ``compression wave'' in $\phi_i$ from initial configuration of
    $\phi_i \in [i-2\pi/N,i+2\pi/N)$, clustered motion about the four initial nodes of the
    compression wave, $\phi_i$ disordered final state. Specifically the values of $t$ are: $t_1 = 0$, $t_2 = 50$, 
    $t_3 = 100$, $t_4 = 200$, 
    $t_5 = 7,000$, $t_6 = 7,200$. \label{fig:positionhists}}
\end{figure}

\begin{figure}[h]
    \centering
    \includegraphics[width=8.0 cm]{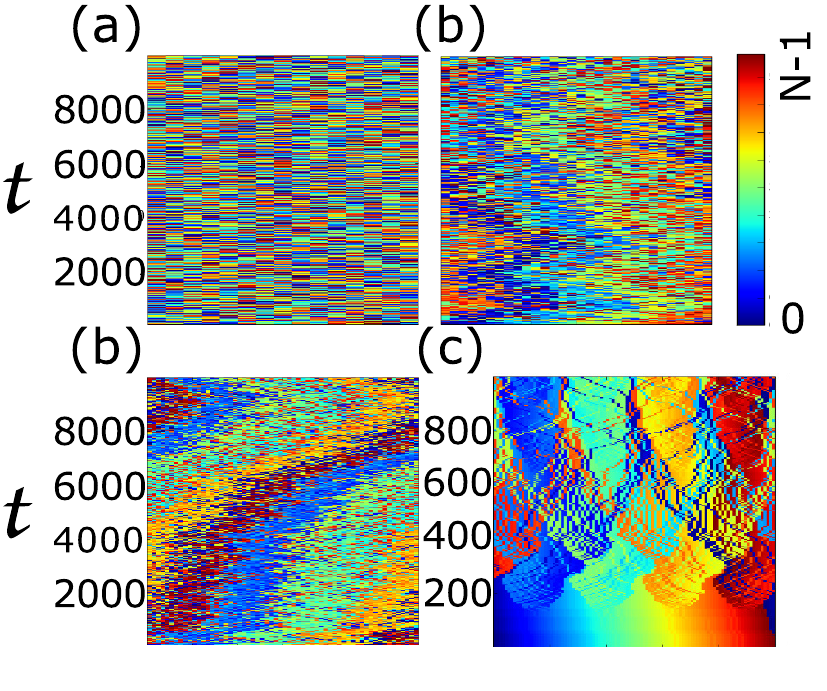}
    \caption{ Each particle is colored from blue, representing the smallest index, to red,
    representing the largest index. The order of the particles indices at a given moment in time 
    is plotted along the horizontal axis. Time increases along the vertical axis.(a) $N=15$. (b)
    $N=30$. (c) $N=60$. (d) $N=100$, where a smaller range of time is shown in order to see the
    mixing of the indices as the angles begin to cluster. 
\label{fig:index_color_phi}}
\end{figure}

Initially the bobs oscillate with an amplitude that depends sinusoidally on their position in $\phi$, 
similar to the previous discussion in the $x$ picture but with four nodes where the $\phi_i$ remain
relatively stationary. Once again this behavior could be thought of as a standing compression wave,
but in $\phi_i$ it has a wave length of $\pi$ whereas in the $x$ picture it had a wavelength of
$2\pi$.
As the system evolves, all $\phi_i$ slowly begin to shift towards the nodes of this standing wave
until there are four clusters of the angles. 
After some time,
the angles begin to re-distribute themselves randomly about the domain. The distribution of $\phi_i$
in these three regimes is summarized in three histograms shown in Fig.~\ref{fig:positionhists}.
Aside from the number of clusters, there are two primary differences between the clustering in $\phi$ and the
clustering in $x$: (i) The clustering in $\phi$ \textit{only} occurs when the angles are
ordered in the method described above, whereas the dynamics in $x$ look identical regardless of the
distribution in $x$, presuming it is somewhat homogeneous about the domain. (ii) The clustering in
$\phi$ is a quasistationary state whereas the clustering in $x$ exists for much longer times
regardless of the system size.
Since the clustering in $\phi$ is quasistationary, a properly prepared system could exist in the
clustered angle state for an arbitrarily long time but only for large $N$. We can view the effect of
increasing $N$ and therefore the lifetimes of the clustered states by observing the order of the
particle index as a function of time. In Fig.~\ref{fig:index_color_phi}(a)-(c),
we show that as $N$ is increased, the time it takes for particles to fully mix increases. This is
shown by plotting the indices on a color scale from 0 (blue) to
$N-1$ (red) along the horizontal axis as time is increased along the vertical axis. In
Fig.~\ref{fig:index_color_phi}(d), we show how the ordering of the particles changes at the very
beginning of clustering for $N=100$. Figure.~\ref{fig:index_color_phi} also shows that the
compression wave is not quasistationary since it quickly reduces to the clustered state regardless of
$N$.

\section{Conclusion}

Though the application of statistical mechanics and thermodynamics to systems with long-range
interactions may not always be appropriate, we find that the canonical partition function improves
our understanding of a system of pendula with long-range interacting bobs. Solving for the canonical
partition function of the Hamiltonian in Eq.~(\ref{eqn:ourH}), we show that the equilibrium behavior
in the $\phi$ coordinate frame is equivalent to the $x$ coordinate frame, i.e. the HMF model. 
As we have argued that the
Hamiltonian in Eq.~(\ref{eqn:ourH}) describes the behavior of the angles of repulsive or attracting pendulum bobs 
(see Fig.~\ref{fig:system}), then the proven equivalence of the canonical partition
function of Eq.~(\ref{eqn:ourH}) and the Hamiltonian mean field model suggests that the Hamiltonian
mean field model sufficiently describes the angles of a system of pendula in equilibrium. Ensemble equivalence
between the microcanonical ensemble and the canonical ensemble is known for the Hamiltonian mean
field model model \cite{campa2009} and because of this, the microcanonical simulations could be used
to approximate equilibrium behavior.  We find numerically that in the case of large
system lengths, $L$, the dynamics of the system in $\phi$ resemble the dynamics of the Hamiltonian
mean field  model, equivalently the behavior of the system in $x$. Therefore for large system sizes
of long pendula in equilibrium, the HMF model describes their dynamics and statistics.  

In this paper we also briefly discuss two particular sets of non-equilibrium results. 
In one case, the system is initialized with small $\phi_i$ so that $x_i$
are distributed relatively evenly throughout the $x$ domain. This initial configuration essentially
gives rise to the ``repulsive'' low temperature HMF model which exhibits interesting non-equilibrium behavior and
is described in great detail by \cite{dauxois2000,barre2001}. In the second case, in which $\phi_i$ are
ordered by their index $i$, we show there is a compression wave in
$\phi$, followed by clustering, and finally a mixed index state displaying no apparent order or
structure. This is in contrast to the dynamics produced by a randomly distributed set of initial
$\phi_i$ which begins and then remains in a random disordered state.
The clustering that can occur in $\phi$ is different from the
clustering in $x$ because it only occurs when the angles are initially ordered and because 
it is quasistationary; the lifetime increases with the number of particles in the system.

\begin{acknowledgments}
We would like to acknowledge the support and generosity of Anand Sharma. This work is partially supported by a
contract from NASA (NNX13AD40A).
\end{acknowledgments}

\bibliography{long_range_refs}

\begin{thebibliography}{21}%
\makeatletter
\providecommand \@ifxundefined [1]{%
 \@ifx{#1\undefined}
}%
\providecommand \@ifnum [1]{%
 \ifnum #1\expandafter \@firstoftwo
 \else \expandafter \@secondoftwo
 \fi
}%
\providecommand \@ifx [1]{%
 \ifx #1\expandafter \@firstoftwo
 \else \expandafter \@secondoftwo
 \fi
}%
\providecommand \natexlab [1]{#1}%
\providecommand \enquote  [1]{``#1''}%
\providecommand \bibnamefont  [1]{#1}%
\providecommand \bibfnamefont [1]{#1}%
\providecommand \citenamefont [1]{#1}%
\providecommand \href@noop [0]{\@secondoftwo}%
\providecommand \href [0]{\begingroup \@sanitize@url \@href}%
\providecommand \@href[1]{\@@startlink{#1}\@@href}%
\providecommand \@@href[1]{\endgroup#1\@@endlink}%
\providecommand \@sanitize@url [0]{\catcode `\\12\catcode `\$12\catcode
  `\&12\catcode `\#12\catcode `\^12\catcode `\_12\catcode `\%12\relax}%
\providecommand \@@startlink[1]{}%
\providecommand \@@endlink[0]{}%
\providecommand \url  [0]{\begingroup\@sanitize@url \@url }%
\providecommand \@url [1]{\endgroup\@href {#1}{\urlprefix }}%
\providecommand \urlprefix  [0]{URL }%
\providecommand \Eprint [0]{\href }%
\providecommand \doibase [0]{http://dx.doi.org/}%
\providecommand \selectlanguage [0]{\@gobble}%
\providecommand \bibinfo  [0]{\@secondoftwo}%
\providecommand \bibfield  [0]{\@secondoftwo}%
\providecommand \translation [1]{[#1]}%
\providecommand \BibitemOpen [0]{}%
\providecommand \bibitemStop [0]{}%
\providecommand \bibitemNoStop [0]{.\EOS\space}%
\providecommand \EOS [0]{\spacefactor3000\relax}%
\providecommand \BibitemShut  [1]{\csname bibitem#1\endcsname}%
\let\auto@bib@innerbib\@empty
%</preamble>
\bibitem [{\citenamefont {Dauxois}\ \emph {et~al.}(2002)\citenamefont
  {Dauxois}, \citenamefont {Ruffo}, \citenamefont {Arimondo},\ and\
  \citenamefont {Wilkens}}]{dynam_therm_intro}%
  \BibitemOpen
  \bibfield  {author} {\bibinfo {author} {\bibfnamefont {T.}~\bibnamefont
  {Dauxois}}, \bibinfo {author} {\bibfnamefont {S.}~\bibnamefont {Ruffo}},
  \bibinfo {author} {\bibfnamefont {E.}~\bibnamefont {Arimondo}}, \ and\
  \bibinfo {author} {\bibfnamefont {M.}~\bibnamefont {Wilkens}},\ }in\ \href
  {\doibase 10.1007/3-540-45835-2_1} {\emph {\bibinfo {booktitle} {Dynamics and
  Thermodynamics of Systems with Long-Range Interactions}}},\ \bibinfo {series}
  {Lecture Notes in Physics}, Vol.\ \bibinfo {volume} {602},\ \bibinfo {editor}
  {edited by\ \bibinfo {editor} {\bibfnamefont {T.}~\bibnamefont {Dauxois}},
  \bibinfo {editor} {\bibfnamefont {S.}~\bibnamefont {Ruffo}}, \bibinfo
  {editor} {\bibfnamefont {E.}~\bibnamefont {Arimondo}}, \ and\ \bibinfo
  {editor} {\bibfnamefont {M.}~\bibnamefont {Wilkens}}}\ (\bibinfo  {publisher}
  {Springer Berlin Heidelberg},\ \bibinfo {year} {2002})\ pp.\ \bibinfo {pages}
  {1--19}\BibitemShut {NoStop}%
\bibitem [{\citenamefont {Kac}\ \emph {et~al.}(1963)\citenamefont {Kac},
  \citenamefont {Uhlenbeck},\ and\ \citenamefont {Hemmer}}]{kac1963}%
  \BibitemOpen
  \bibfield  {author} {\bibinfo {author} {\bibfnamefont {M.}~\bibnamefont
  {Kac}}, \bibinfo {author} {\bibfnamefont {G.~E.}\ \bibnamefont {Uhlenbeck}},
  \ and\ \bibinfo {author} {\bibfnamefont {P.~C.}\ \bibnamefont {Hemmer}},\
  }\href {\doibase http://dx.doi.org/10.1063/1.1703946} {\bibfield  {journal}
  {\bibinfo  {journal} {Journal of Mathematical Physics}\ }\textbf {\bibinfo
  {volume} {4}},\ \bibinfo {pages} {216} (\bibinfo {year} {1963})}\BibitemShut
  {NoStop}%
\bibitem [{\citenamefont {Sire}\ and\ \citenamefont
  {Chavanis}(2002)}]{sire2002}%
  \BibitemOpen
  \bibfield  {author} {\bibinfo {author} {\bibfnamefont {C.}~\bibnamefont
  {Sire}}\ and\ \bibinfo {author} {\bibfnamefont {P.-H.}\ \bibnamefont
  {Chavanis}},\ }\href {\doibase 10.1103/PhysRevE.66.046133} {\bibfield
  {journal} {\bibinfo  {journal} {Physical Review E}\ }\textbf {\bibinfo
  {volume} {66}},\ \bibinfo {pages} {046133} (\bibinfo {year}
  {2002})}\BibitemShut {NoStop}%
\bibitem [{\citenamefont {Antoniazzi}\ \emph {et~al.}(2007)\citenamefont
  {Antoniazzi}, \citenamefont {Fanelli}, \citenamefont {Barr\'{e}},
  \citenamefont {Chavanis}, \citenamefont {Dauxois},\ and\ \citenamefont
  {Ruffo}}]{antoniazzi2007}%
  \BibitemOpen
  \bibfield  {author} {\bibinfo {author} {\bibfnamefont {A.}~\bibnamefont
  {Antoniazzi}}, \bibinfo {author} {\bibfnamefont {D.}~\bibnamefont {Fanelli}},
  \bibinfo {author} {\bibfnamefont {J.}~\bibnamefont {Barr\'{e}}}, \bibinfo
  {author} {\bibfnamefont {P.-H.}\ \bibnamefont {Chavanis}}, \bibinfo {author}
  {\bibfnamefont {T.}~\bibnamefont {Dauxois}}, \ and\ \bibinfo {author}
  {\bibfnamefont {S.}~\bibnamefont {Ruffo}},\ }\href {\doibase
  10.1103/PhysRevE.75.011112} {\bibfield  {journal} {\bibinfo  {journal}
  {Physical Review E}\ }\textbf {\bibinfo {volume} {75}},\ \bibinfo {pages}
  {011112} (\bibinfo {year} {2007})}\BibitemShut {NoStop}%
\bibitem [{\citenamefont {Christodoulidi}\ \emph {et~al.}(2014)\citenamefont
  {Christodoulidi}, \citenamefont {Tsallis},\ and\ \citenamefont
  {Bountis}}]{christodoulidi2014}%
  \BibitemOpen
  \bibfield  {author} {\bibinfo {author} {\bibfnamefont {H.}~\bibnamefont
  {Christodoulidi}}, \bibinfo {author} {\bibfnamefont {C.}~\bibnamefont
  {Tsallis}}, \ and\ \bibinfo {author} {\bibfnamefont {T.}~\bibnamefont
  {Bountis}},\ }\href {http://stacks.iop.org/0295-5075/108/i=4/a=40006}
  {\bibfield  {journal} {\bibinfo  {journal} {EPL (Europhysics Letters)}\
  }\textbf {\bibinfo {volume} {108}},\ \bibinfo {pages} {40006} (\bibinfo
  {year} {2014})}\BibitemShut {NoStop}%
\bibitem [{\citenamefont {Antoni}\ and\ \citenamefont
  {Ruffo}(1995)}]{antoni1995}%
  \BibitemOpen
  \bibfield  {author} {\bibinfo {author} {\bibfnamefont {M.}~\bibnamefont
  {Antoni}}\ and\ \bibinfo {author} {\bibfnamefont {S.}~\bibnamefont {Ruffo}},\
  }\href {http://pre.aps.org/abstract/PRE/v52/i3/p2361\_1} {\bibfield
  {journal} {\bibinfo  {journal} {Physical Review E}\ }\textbf {\bibinfo
  {volume} {52}} (\bibinfo {year} {1995})}\BibitemShut {NoStop}%
\bibitem [{\citenamefont {Inouye}\ \emph {et~al.}(1998)\citenamefont {Inouye},
  \citenamefont {Andrews},\ and\ \citenamefont {Stenger}}]{inouye1998}%
  \BibitemOpen
  \bibfield  {author} {\bibinfo {author} {\bibfnamefont {S.}~\bibnamefont
  {Inouye}}, \bibinfo {author} {\bibfnamefont {M.}~\bibnamefont {Andrews}}, \
  and\ \bibinfo {author} {\bibfnamefont {J.}~\bibnamefont {Stenger}},\ }\href
  {http://www.nature.com/nature/journal/v392/n6672/abs/392151a0.html}
  {\bibfield  {journal} {\bibinfo  {journal} {Nature}\ }\textbf {\bibinfo
  {volume} {392}},\ \bibinfo {pages} {151} (\bibinfo {year}
  {1998})}\BibitemShut {NoStop}%
\bibitem [{\citenamefont {O'dell}\ \emph {et~al.}(2000)\citenamefont {O'dell},
  \citenamefont {Giovanazzi}, \citenamefont {Kurizki},\ and\ \citenamefont
  {Akulin}}]{odell2000}%
  \BibitemOpen
  \bibfield  {author} {\bibinfo {author} {\bibfnamefont {D.}~\bibnamefont
  {O'dell}}, \bibinfo {author} {\bibfnamefont {S.}~\bibnamefont {Giovanazzi}},
  \bibinfo {author} {\bibfnamefont {G.}~\bibnamefont {Kurizki}}, \ and\
  \bibinfo {author} {\bibfnamefont {V.}~\bibnamefont {Akulin}},\ }\href
  {http://journals.aps.org/prl/abstract/10.1103/PhysRevLett.84.5687} {\bibfield
   {journal} {\bibinfo  {journal} {Physical Review Letters}\ ,\ \bibinfo
  {pages} {5687}} (\bibinfo {year} {2000})}\BibitemShut {NoStop}%
\bibitem [{\citenamefont {Campa}\ \emph {et~al.}(2009)\citenamefont {Campa},
  \citenamefont {Dauxois},\ and\ \citenamefont {Ruffo}}]{campa2009}%
  \BibitemOpen
  \bibfield  {author} {\bibinfo {author} {\bibfnamefont {A.}~\bibnamefont
  {Campa}}, \bibinfo {author} {\bibfnamefont {T.}~\bibnamefont {Dauxois}}, \
  and\ \bibinfo {author} {\bibfnamefont {S.}~\bibnamefont {Ruffo}},\ }\href
  {\doibase 10.1016/j.physrep.2009.07.001} {\bibfield  {journal} {\bibinfo
  {journal} {Physics Reports}\ }\textbf {\bibinfo {volume} {480}},\ \bibinfo
  {pages} {57} (\bibinfo {year} {2009})}\BibitemShut {NoStop}%
\bibitem [{\citenamefont {Campa}\ \emph {et~al.}(2000)\citenamefont {Campa},
  \citenamefont {Giansanti},\ and\ \citenamefont {Moroni}}]{campa2000}%
  \BibitemOpen
  \bibfield  {author} {\bibinfo {author} {\bibfnamefont {A.}~\bibnamefont
  {Campa}}, \bibinfo {author} {\bibfnamefont {A.}~\bibnamefont {Giansanti}}, \
  and\ \bibinfo {author} {\bibfnamefont {D.}~\bibnamefont {Moroni}},\ }\href
  {\doibase 10.1103/PhysRevE.62.303} {\bibfield  {journal} {\bibinfo  {journal}
  {Phys. Rev. E}\ }\textbf {\bibinfo {volume} {62}},\ \bibinfo {pages} {303}
  (\bibinfo {year} {2000})}\BibitemShut {NoStop}%
\bibitem [{\citenamefont {Campa}\ \emph {et~al.}(2003)\citenamefont {Campa},
  \citenamefont {Giansanti},\ and\ \citenamefont {Moroni}}]{campa2003}%
  \BibitemOpen
  \bibfield  {author} {\bibinfo {author} {\bibfnamefont {A.}~\bibnamefont
  {Campa}}, \bibinfo {author} {\bibfnamefont {A.}~\bibnamefont {Giansanti}}, \
  and\ \bibinfo {author} {\bibfnamefont {D.}~\bibnamefont {Moroni}},\ }\href
  {\doibase 10.1088/0305-4470/36/25/301} {\bibfield  {journal} {\bibinfo
  {journal} {Journal of Physics A: Mathematical and General}\ }\textbf
  {\bibinfo {volume} {36}},\ \bibinfo {pages} {6897} (\bibinfo {year}
  {2003})}\BibitemShut {NoStop}%
\bibitem [{\citenamefont {Anteneodo}\ and\ \citenamefont
  {Tsallis}(1998)}]{anteneodo1998}%
  \BibitemOpen
  \bibfield  {author} {\bibinfo {author} {\bibfnamefont {C.}~\bibnamefont
  {Anteneodo}}\ and\ \bibinfo {author} {\bibfnamefont {C.}~\bibnamefont
  {Tsallis}},\ }\href {\doibase 10.1103/PhysRevLett.80.5313} {\bibfield
  {journal} {\bibinfo  {journal} {Phys. Rev. Lett.}\ }\textbf {\bibinfo
  {volume} {80}},\ \bibinfo {pages} {5313} (\bibinfo {year}
  {1998})}\BibitemShut {NoStop}%
\bibitem [{\citenamefont {Tamarit}\ and\ \citenamefont
  {Anteneodo}(2000)}]{tamarit2000}%
  \BibitemOpen
  \bibfield  {author} {\bibinfo {author} {\bibfnamefont {F.}~\bibnamefont
  {Tamarit}}\ and\ \bibinfo {author} {\bibfnamefont {C.}~\bibnamefont
  {Anteneodo}},\ }\href {\doibase 10.1103/PhysRevLett.84.208} {\bibfield
  {journal} {\bibinfo  {journal} {Phys. Rev. Lett.}\ }\textbf {\bibinfo
  {volume} {84}},\ \bibinfo {pages} {208} (\bibinfo {year} {2000})}\BibitemShut
  {NoStop}%
\bibitem [{\citenamefont {Cirto}\ \emph {et~al.}(2014)\citenamefont {Cirto},
  \citenamefont {Assis},\ and\ \citenamefont {Tsallis}}]{cirto2014}%
  \BibitemOpen
  \bibfield  {author} {\bibinfo {author} {\bibfnamefont {L.~J.}\ \bibnamefont
  {Cirto}}, \bibinfo {author} {\bibfnamefont {V.~R.}\ \bibnamefont {Assis}}, \
  and\ \bibinfo {author} {\bibfnamefont {C.}~\bibnamefont {Tsallis}},\ }\href
  {\doibase http://dx.doi.org/10.1016/j.physa.2013.09.002} {\bibfield
  {journal} {\bibinfo  {journal} {Physica A: Statistical Mechanics and its
  Applications}\ }\textbf {\bibinfo {volume} {393}},\ \bibinfo {pages} {286 }
  (\bibinfo {year} {2014})}\BibitemShut {NoStop}%
\bibitem [{\citenamefont {Levin}\ \emph {et~al.}(2014)\citenamefont {Levin},
  \citenamefont {Pakter}, \citenamefont {Rizzato}, \citenamefont {Teles},\ and\
  \citenamefont {Benetti}}]{levin2014}%
  \BibitemOpen
  \bibfield  {author} {\bibinfo {author} {\bibfnamefont {Y.}~\bibnamefont
  {Levin}}, \bibinfo {author} {\bibfnamefont {R.}~\bibnamefont {Pakter}},
  \bibinfo {author} {\bibfnamefont {F.~B.}\ \bibnamefont {Rizzato}}, \bibinfo
  {author} {\bibfnamefont {T.~N.}\ \bibnamefont {Teles}}, \ and\ \bibinfo
  {author} {\bibfnamefont {F.~P.}\ \bibnamefont {Benetti}},\ }\href {\doibase
  http://dx.doi.org/10.1016/j.physrep.2013.10.001} {\bibfield  {journal}
  {\bibinfo  {journal} {Physics Reports}\ }\textbf {\bibinfo {volume} {535}},\
  \bibinfo {pages} {1 } (\bibinfo {year} {2014})},\ \bibinfo {note}
  {nonequilibrium statistical mechanics of systems with long-range
  interactions}\BibitemShut {NoStop}%
\bibitem [{\citenamefont {Teles}\ \emph {et~al.}(2012)\citenamefont {Teles},
  \citenamefont {Benetti}, \citenamefont {Pakter},\ and\ \citenamefont
  {Levin}}]{teles2012}%
  \BibitemOpen
  \bibfield  {author} {\bibinfo {author} {\bibfnamefont {T.~N.}\ \bibnamefont
  {Teles}}, \bibinfo {author} {\bibfnamefont {F.~P. D.~C.}\ \bibnamefont
  {Benetti}}, \bibinfo {author} {\bibfnamefont {R.}~\bibnamefont {Pakter}}, \
  and\ \bibinfo {author} {\bibfnamefont {Y.}~\bibnamefont {Levin}},\ }\href
  {\doibase 10.1103/PhysRevLett.109.230601} {\bibfield  {journal} {\bibinfo
  {journal} {Physical Review Letters}\ }\textbf {\bibinfo {volume} {109}},\
  \bibinfo {pages} {230601} (\bibinfo {year} {2012})}\BibitemShut {NoStop}%
\bibitem [{\citenamefont {Altland}\ and\ \citenamefont
  {Simons}(2006)}]{altland2006}%
  \BibitemOpen
  \bibfield  {author} {\bibinfo {author} {\bibfnamefont {A.}~\bibnamefont
  {Altland}}\ and\ \bibinfo {author} {\bibfnamefont {B.}~\bibnamefont
  {Simons}},\ }\href {http://books.google.com/books?id=0KMkfAMe3JkC} {\emph
  {\bibinfo {title} {Condensed Matter Field Theory}}}\ (\bibinfo  {publisher}
  {Cambridge University Press},\ \bibinfo {year} {2006})\BibitemShut {NoStop}%
\bibitem [{\citenamefont {Miritello}\ \emph {et~al.}(2009)\citenamefont
  {Miritello}, \citenamefont {Pluchino},\ and\ \citenamefont
  {Rapisarda}}]{miritello2009}%
  \BibitemOpen
  \bibfield  {author} {\bibinfo {author} {\bibfnamefont {G.}~\bibnamefont
  {Miritello}}, \bibinfo {author} {\bibfnamefont {a.}~\bibnamefont {Pluchino}},
  \ and\ \bibinfo {author} {\bibfnamefont {a.}~\bibnamefont {Rapisarda}},\
  }\href {\doibase 10.1209/0295-5075/85/10007} {\bibfield  {journal} {\bibinfo
  {journal} {EPL (Europhysics Letters)}\ }\textbf {\bibinfo {volume} {85}},\
  \bibinfo {pages} {10007} (\bibinfo {year} {2009})}\BibitemShut {NoStop}%
\bibitem [{\citenamefont {Barré}\ \emph {et~al.}(2001)\citenamefont {Barré},
  \citenamefont {Dauxois},\ and\ \citenamefont {Ruffo}}]{barre2001}%
  \BibitemOpen
  \bibfield  {author} {\bibinfo {author} {\bibfnamefont {J.}~\bibnamefont
  {Barré}}, \bibinfo {author} {\bibfnamefont {T.}~\bibnamefont {Dauxois}}, \
  and\ \bibinfo {author} {\bibfnamefont {S.}~\bibnamefont {Ruffo}},\ }\href
  {\doibase http://dx.doi.org/10.1016/S0378-4371(01)00084-X} {\bibfield
  {journal} {\bibinfo  {journal} {Physica A: Statistical Mechanics and its
  Applications}\ }\textbf {\bibinfo {volume} {295}},\ \bibinfo {pages} {254 }
  (\bibinfo {year} {2001})},\ \bibinfo {note} {proceedings of the {IUPAP}
  International Conference on New Trends in the Fractal Aspects of Complex
  Systems}\BibitemShut {NoStop}%
\bibitem [{\citenamefont {Leyvraz}\ \emph {et~al.}(2002)\citenamefont
  {Leyvraz}, \citenamefont {Firpo},\ and\ \citenamefont {Ruffo}}]{leyvraz2002}%
  \BibitemOpen
  \bibfield  {author} {\bibinfo {author} {\bibfnamefont {F.}~\bibnamefont
  {Leyvraz}}, \bibinfo {author} {\bibfnamefont {M.-C.}\ \bibnamefont {Firpo}},
  \ and\ \bibinfo {author} {\bibfnamefont {S.}~\bibnamefont {Ruffo}},\ }\href
  {http://stacks.iop.org/0305-4470/35/i=20/a=303} {\bibfield  {journal}
  {\bibinfo  {journal} {Journal of Physics A: Mathematical and General}\
  }\textbf {\bibinfo {volume} {35}},\ \bibinfo {pages} {4413} (\bibinfo {year}
  {2002})}\BibitemShut {NoStop}%
\bibitem [{\citenamefont {Dauxois}\ \emph {et~al.}(2000)\citenamefont
  {Dauxois}, \citenamefont {Holdsworth},\ and\ \citenamefont
  {Ruffo}}]{dauxois2000}%
  \BibitemOpen
  \bibfield  {author} {\bibinfo {author} {\bibfnamefont {T.}~\bibnamefont
  {Dauxois}}, \bibinfo {author} {\bibfnamefont {P.}~\bibnamefont {Holdsworth}},
  \ and\ \bibinfo {author} {\bibfnamefont {S.}~\bibnamefont {Ruffo}},\ }\href
  {http://link.springer.com/article/10.1007/s100510070183} {\bibfield
  {journal} {\bibinfo  {journal} {The European Physical Journal B- \ldots}\
  }\textbf {\bibinfo {volume} {16}},\ \bibinfo {pages} {659} (\bibinfo {year}
  {2000})}\BibitemShut {NoStop}%
\end{thebibliography}%

\end{document}